\documentclass[11pt,twoside]{article}
\usepackage{asp2010}

\resetcounters

\begin{document}

\title{Simulations of chromospheric heating by ambipolar diffusion}

\author{E. Khomenko$^{1,2}$ and M. Collados$^{1,2}$
\affil{$^1$Instituto de Astrof\'{\i}sica de Canarias, 38205, C/ V\'{\i}a L{\'a}ctea, s/n, La Laguna, Tenerife, Spain}
\affil{$^2$Departamento de Astrof\'{\i}sica, Universidad de La Laguna, 38205, La Laguna, Tenerife, Spain}}

\begin{abstract}
We propose a mechanism for efficient heating of the solar chromosphere, based
on non-ideal plasma effects. Three ingredients are needed for the work of
this mechanism: (1) presence of neutral atoms; (2) presence of a
non-potential magnetic field; (3) decrease of the collisional coupling of the
plasma. Due to decrease of collisional coupling, a net relative motion
appears between the neutral and ionized components, usually referred to as
``ambipolar diffusion''. This results in a significant enhancement of current
dissipation as compared to the classical MHD case. We propose that the
current dissipation in this situation is able to provide enough energy to
heat the chromosphere by several kK on the time scale of minutes, or even
seconds. In this paper, we show that this energy supply might be sufficient
to balance the radiative energy losses of the chromosphere.
\end{abstract}

\section{Introduction}

\begin{figure}[t]
\center
\plotfiddle{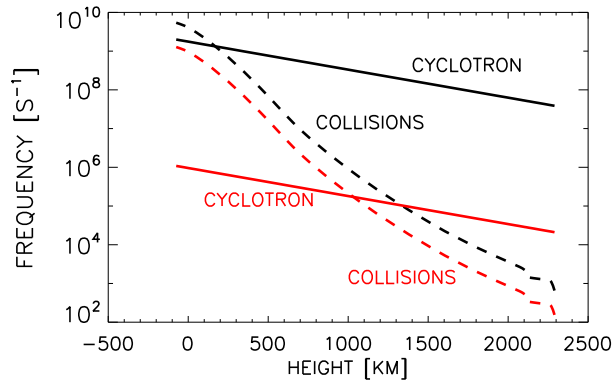}{5cm}{0.}{70.}{70.}{-100}{0}
\caption{{\footnotesize Estimation of the collisional and cyclotron
frequencies for electrons (black lines) and hydrogen ions (red lines) in the
photosphere and chromosphere of the Sun. The magnetic field strength is
assumed to vary with height as $B(z)=B_0\exp{(-z/H_B)}$, with $B_0=100$ G and
$H_B=600$ km.}}
\label{fig:colis}
\end{figure}

The degree of plasma ionization in the lower solar atmosphere $-$ photosphere
and chromosphere $-$ is very small. Using VAL-C model atmosphere as a
reference \citep[][]{1981ApJS...45..635V}, the abundance of ionized atoms, relative to
neutral atoms, is as low as $10^{-4}$ at heights of temperature minimum, and
it remains always well below unity even at larger heights. In addition to
that, the collisional coupling of the plasma becomes less important with
height. A simple calculation reveals that cyclotron frequency of hydrogen
ions may exceed the collisional frequency already in the lower photosphere,
for values of magnetic field strength expected for the quiet photosphere, see Figure
\ref{fig:colis}.

These two factors may lead to a break of the assumption underlying
magnetohydrodynamics (MHD) and lead to new effects, not taken into account in
the classical approach, as ambipolar diffusion. In astrophysics, ambipolar
(or neutral) diffusion usually refers to the decoupling of neutral and
charged components. Ambipolar diffusion causes the magnetic field to diffuse
through neutral gas due to collisions between neutrals and charged particles,
the latter being frozen-in into the magnetic field.

There is an increasing number of evidences for the importance of deviations
from MHD in different situations. The presence of neutral atoms in partially
ionized plasmas significantly affects wave excitation and propagation
\citep{2003SoPh..214..241K, 2004A&A...422.1073K, 2006AdSpR..37..447K, 2007A&A...461..731F,
2008A&A...478..553V, 2009ApJ...699.1553S, 2010A&A...512A..28S,2011A&A...529A..82Z}. It is also important for
magnetic reconnection \citep{1994ApJ...427L..91B,
1995ApJ...448..734B, 2006ApJ...642.1236S, 2008A&A...486..569S, 2009ApJ...691L..45S}.
Non-ideal plasma effects can modify the equilibrium balance of photospheric
flux tubes \citep{2002Ap&SS.279..389K} and chromospheric structures, such
as prominences \citep{2009ApJ...705.1183A, 2002ApJ...577..464G}. Another phenomenon potentially
affected by non-ideal plasma effects is magnetic flux emergence
\citep{2006A&A...450..805L, 2007ApJ...666..541A}. Despite this increasing evidence, we are
far from a complete understanding of the influence of these effects.

In this paper, we continue the investigation started in
\citet{Khomenko+Collados2012}. There, we studied the consequences of the
ambipolar diffusion into the heating of the magnetized solar chromosphere. In
the presence of neutrals, the ambipolar (or neutral) diffusion is orders of
magnitude larger than the classical Ohmic diffusion, leading to efficient
Joule dissipation of electric currents. Our calculations have demonstrated
that just by existing relatively weak (10--40 G), non-force-free magnetic
fields, the chromospheric layers above 1000 km can be efficiently heated by
current dissipation reaching an increase of temperature of 1-2 kK in a time
interval of minutes. The work of ambipolar diffusion would stop when all the
atoms become ionized or when the magnetic field becomes force-free ($\vec{J}
\parallel \vec{B}$).
We proposed that this heating mechanism may be efficient enough to balance
the radiative losses of the chromosphere
\citep[see][]{Khomenko+Collados2012}.
%
Here we explore this possibility by including the radiative damping in our
initial calculations.

\section{Equations and numerical solution}

We solve numerically the quasi-MHD equations of conservation of mass,
momentum, internal energy, and the induction equation
\citep{Khomenko+Collados2012}.

\begin{eqnarray} \label{eq:system}
\frac{\partial \rho}{\partial t} + \vec{\nabla}\left(\rho\vec{u}\right) = 0 \\ \nonumber
\rho\frac{D\vec{u}}{D t} = \vec{J}\times\vec{B} + \rho\vec{g} - \vec{\nabla}p \\ \nonumber
\frac{1}{(\gamma - 1)}\frac{D p}{D t} + \frac{\gamma}{(\gamma - 1)} p\vec{\nabla}\vec{u} = \eta\mu_0\vec{J^2} + \eta_A\mu_0 \vec{J_{\bot}^2} +Q_{\rm rad}\\ \nonumber
\frac{\partial\vec{B}}{\partial t} = \vec{\nabla}\times \left[ (\vec{u}\times\vec{B}) - \eta\mu_0\vec{J} - \eta_A\mu_0\vec{J_{\bot}} \right]
\end{eqnarray}
where the following definitions are used:
\begin{equation}
\rho = \sum_{\alpha=n,i,e}\rho_{\alpha};\,\,\,
\vec{u} = {1 \over \rho} \sum_{\alpha=n,i,e}(\rho_{\alpha}\vec{u}_{\alpha});\,\,\,
\vec{J} = en_e(\vec{u}_i - \vec{u}_e);\,\,\,
p = \sum_{\alpha=n,i,e}p_{\alpha}
\end{equation}
and $\vec{J}_{\perp}$ is the component of the current perpendicular to the
magnetic field. These equations are produced by summing up the equations for
three different species (electrons ($e$), hydrogen ions ($i$) and neutral
hydrogen ($n$)). When deriving them we neglected the non-diagonal components of
the pressure tensor and assumed that the diffusion velocities
$\vec{w}_{\alpha}=\vec{u} - \vec{u}_{\alpha}$ ($\alpha=e,i,n$) are small,
neglecting terms containing $w_{\alpha}^2$. In the Ohm's law we neglected the
time variation of relative ion-neutral velocity $\vec{u}_i - \vec{u}_e$, the
effects on the currents by partial pressure gradients of the three species,
and the gravity force acting on electrons. The Hall term of the Ohm's law does
not appear in the energy equation and, consequently, has no impact on the
thermal evolution of the system. For consistency, we removed the Hall term
from the induction equation as well. The Ohmic and ambipolar diffusion
coefficients are equal to:
\begin{equation}
\label{eq:etas}
\eta = \frac{m_e(\nu_{ei} + \nu_{en})}{e^2 n_e \mu_0};\,\,\,
\eta_A = \frac{(\rho_n/\rho)^2|B|^2}{(\rho_i\nu_{in} + \rho_e\nu_{en})\mu_0}
\end{equation}
and the collisional frequencies are:
\begin{equation}
\nu_{in}=n_{n}\sqrt{\frac{8 k_B T}{\pi m_{in}}}\sigma_{in};\,\,\,
\nu_{en}=n_{n}\sqrt{\frac{8 k_B T}{\pi m_{en}}}\sigma_{en};\,\,\,
\nu_{ei}= \frac{n_e e^4 \Lambda}{3 m_e^2 \epsilon_0^2} \left(\frac{m_e}{2\pi k_B T} \right)^{3/2} \label{eq:nus}
\end{equation}
where $m_{in}=m_i m_n/(m_i + m_n)$ and $m_{en}=m_e m_n/(m_e + m_n)$. The
respective cross sections are $\sigma_{in}=5\times10^{-19}$ m$^2$ and
$\sigma_{en}=10^{-19}$ m$^2$. $\Lambda$ is the Coulomb logarithm.

After subtracting the equilibrium conditions, these equation are solved by
means of our code {\sc mancha} \citep{2008SoPh..251..589K, 2010ApJ...719..357F,
Khomenko+Collados2012} with the inclusion of the physical ohmic and ambipolar
diffusion terms in the equation of energy conservation and in the induction
equation. As our code propagates non-linear perturbations to the magneto-static
equilibrium, we treat the diffusion terms as perturbations. The radiative
losses term $Q_{\rm rad}$ in the energy equation calculated from the solution
of Radiative Transfer Equation in Local Thermodynamic Equilibrium (LTE) and
grey approximation for the opacity dependence on wavelength $\lambda$.
\begin{equation}
Q_{\rm rad} = -\int_{\lambda}(\vec{\nabla}\vec{F}_{\lambda})d\lambda;\,\,\, \vec{F}_{\lambda}=\int_{4\pi}{I_{\lambda}(\vec{\mu})\vec{\mu}d\Omega}
\end{equation}
where $\vec{F}_{\lambda}$ is radiative energy flux, $I_{\lambda}$ is specific
intensity, $\vec{\mu}$ marks the direction and $d\Omega$ is a solid angle.
Following the philosophy of our code, we also perturb the radiative cooling
term, $\Delta Q_{\rm rad}=Q_{\rm rad} - Q_{\rm rad}(0)$, the zero-order term
$Q_{\rm rad}(0)$ being eliminated from the equations as a part of equilibrium
condition.

The equations are solved in two spatial dimensions, though the vector
quantities are allowed to have three dimensions (2.5D approximation). As the
temperature and the ionization state vary with time, we recalculate the
ionization balance of the atmosphere at each time step, assuming LTE (Saha
equations). We then update the neutral fraction, $\rho_n/\rho$, needed for
the calculation of the ambipolar diffusion coefficient (Eq.~\ref{eq:etas}).

\begin{figure*}[t]
\center
\plotfiddle{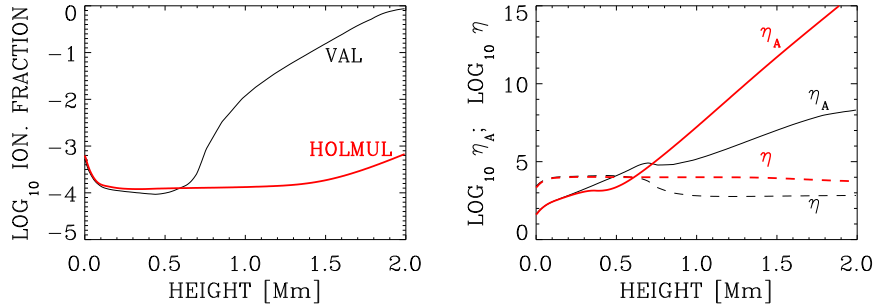}{3.5cm}{0.}{60.}{60.}{-170}{-10}
\caption{{\footnotesize Ionization fraction $\rho_e/\rho$ (left) and
diffusion coefficients $\eta$, $\eta_A$ in m$^2$ s$^{-1}$ (right) as a
function of height in the two flux tube models. Black thin lines: VAL-based
tube; red thick lines: HOLMUL-based tube. }} \label{fig:ft2initial}
\end{figure*}

\section{Flux tube model}

In quiet regions of the solar chromosphere, diagnostic tools based on the
Hanle effect point to magnetic field strengths of the order of tens Gauss
\citep{2005ApJ...619L.191T, 2010ApJ...708.1579C, 2010ApJ...711L.133S}. To simulate
such non-active chromospheric conditions, we used a 2nd-order thin magnetic
flux tube model as initial atmosphere \citep{1986A&A...154..231P,
2008SoPh..251..589K}. The model represents a horizontally infinite series of
flux tubes that merge at some height in the chromosphere, preventing them
from excessive opening with height. This magnetic field configuration is
non-force-free. Here we used two flux tube models, the main difference
between them is their (horizontally homogeneous) temperature structure. One
has the vertical temperature structure of VAL-C \citep{1981ApJS...45..635V}, and the other has
that of HOLMUL \citep{1974SoPh...39...19H}. The magnetic field strength is similar in
both cases, decreasing with height, from about 800 G in the photosphere to 35
G in the chromosphere. In the rest of the text, we will refer to these models
as VAL-based and HOLMUL-based flux tubes.

While the VAL-C model atmosphere includes the chromospheric temperature increase,
the temperature in the HOLMUL model monotonically decreases with height
dictated by the conditions of radiative equilibrium. This peculiarity
determines the ionization fraction and value of the ambipolar diffusion
coefficients in the models (see Fig.~\ref{fig:ft2initial}).
In the HOLMUL-based tube, the ionization fraction value does not exceed $10^{-3}$,
even in the chromosphere. The values of the Ohmic diffusion coefficient $\eta$
(see Eq.~\ref{eq:etas}) are similar in both models. The ambipolar diffusion
$\eta_A$ is orders of magnitude larger than $\eta$ already above 1 Mm height.
Due to the much lower ionization fraction in the HOLMUL-based model, the
value of $\eta_A$ there significantly exceeds that of the VAL-based model.
Such values of $\eta_A$ imply important current dissipation on very short
time scales and a much quicker dissipation is expected in the cooler
HOLMUL-based model.

\begin{figure*}[t]
\center
\plotfiddle{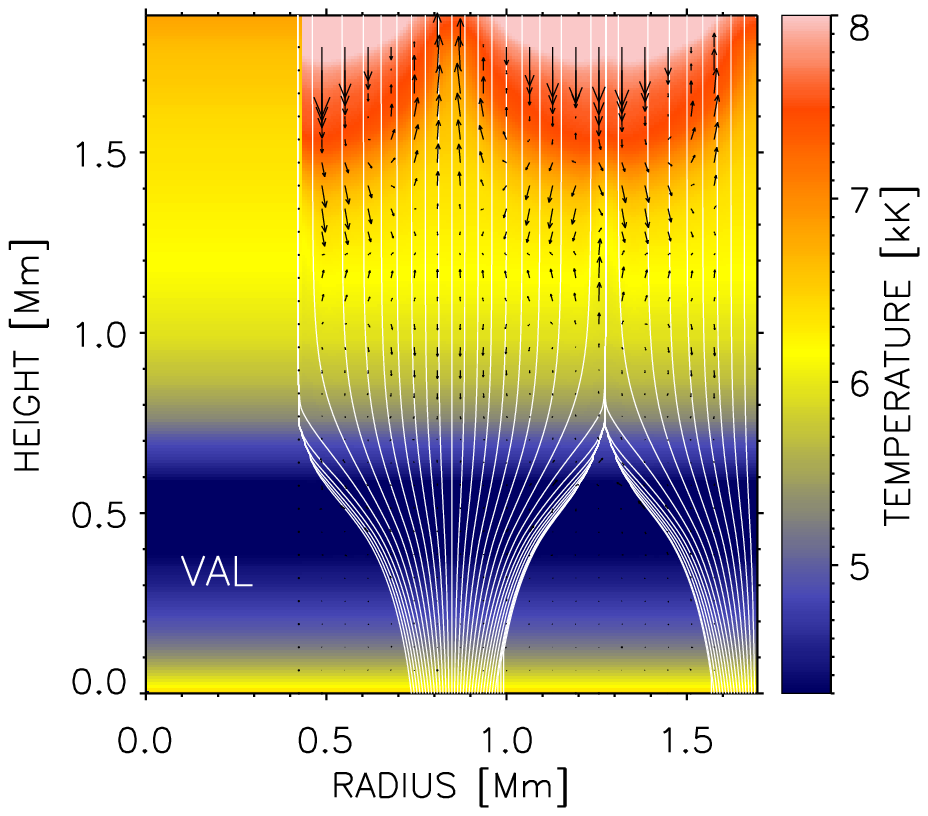}{8.7cm}{0.}{70.}{70.}{-190}{90}
\plotfiddle{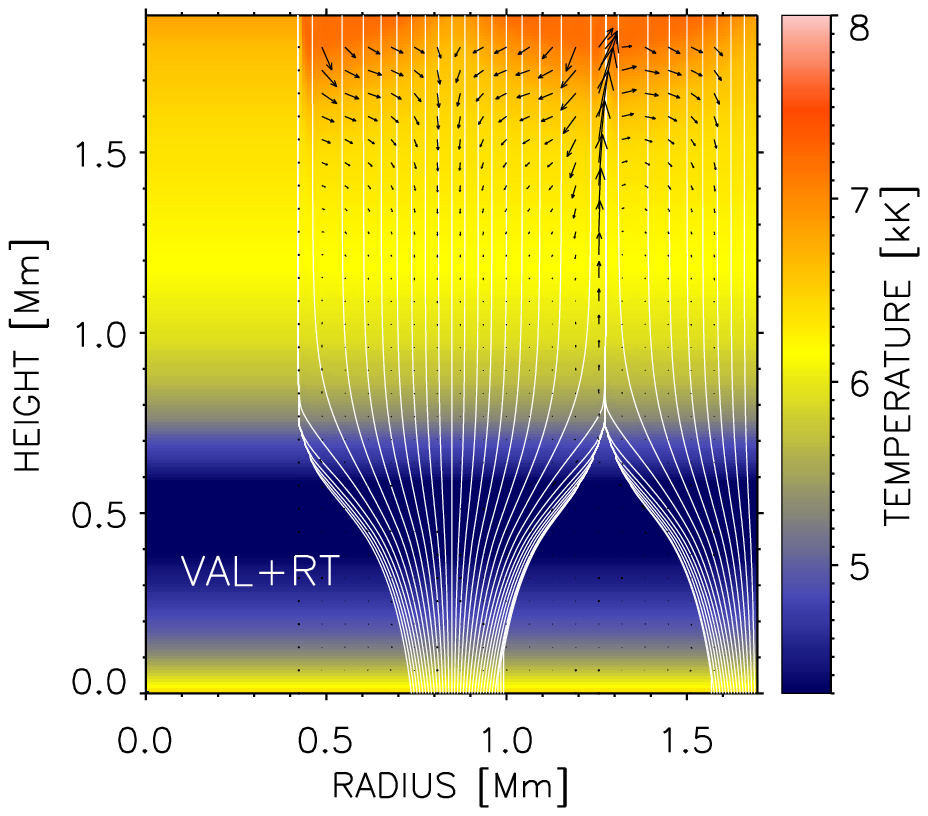}{0cm}{0.}{70.}{70.}{10}{115}
\plotfiddle{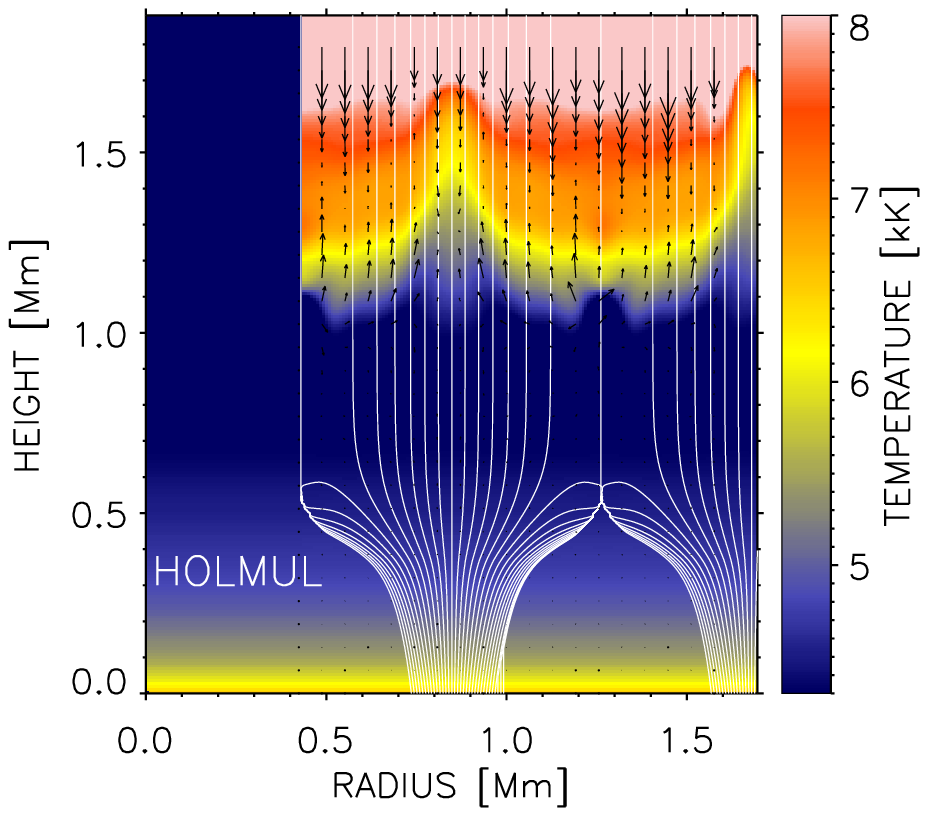}{0cm}{0.}{70.}{70.}{-190}{-30}
\plotfiddle{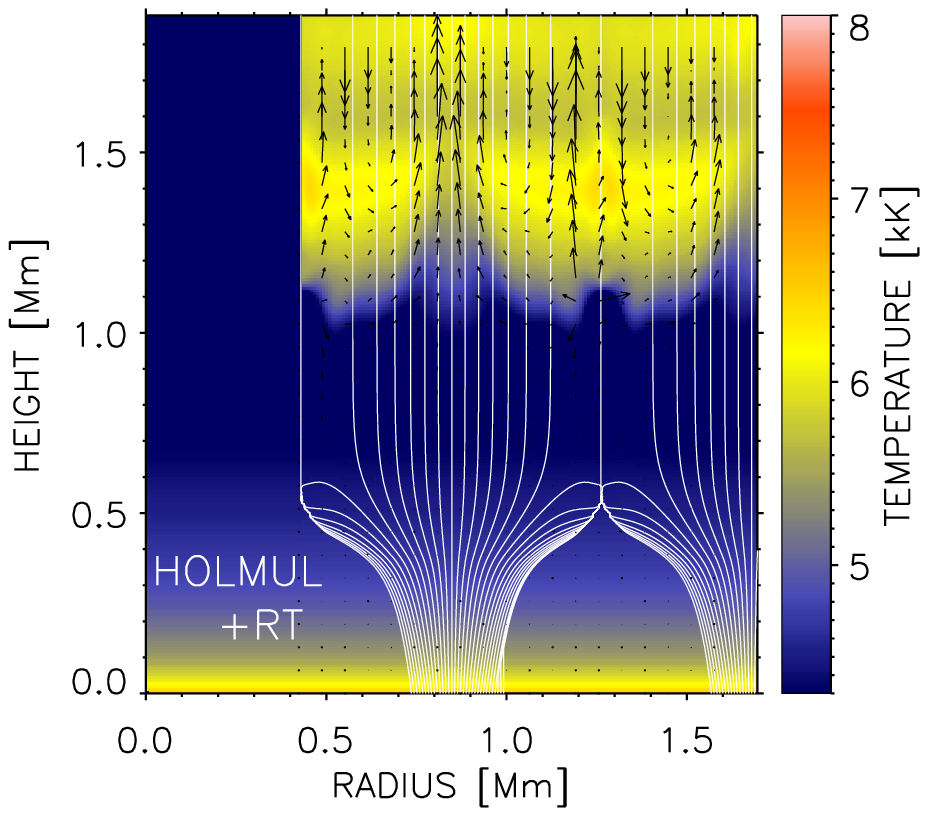}{0cm}{0.}{70.}{70.}{10}{-5}
\caption{{\footnotesize Simulation snapshots after 800 s of evolution. {\it
Top left}: VAL-C based flux tube with $Q_{\rm rad}=0$; {\it top right:} same
but $Q_{\rm rad} \neq 0$; {\it bottom left:} HOLMUL-based flux tube with
$Q_{\rm rad}=0$; {\it bottom right:} same, but $Q_{\rm rad} \neq 0$. The
background color is temperature, the scale is the same for all panels.
Vertical white lines are magnetic field lines. Arrows show the velocity
field. For better visual comparison we keep unchanged the temperature
structure from 0 to 0.42 Mm, though the variations are present in the
simulations.}} \label{fig:ft2evol}
\end{figure*}

\begin{figure*}[t]
\center
\plotfiddle{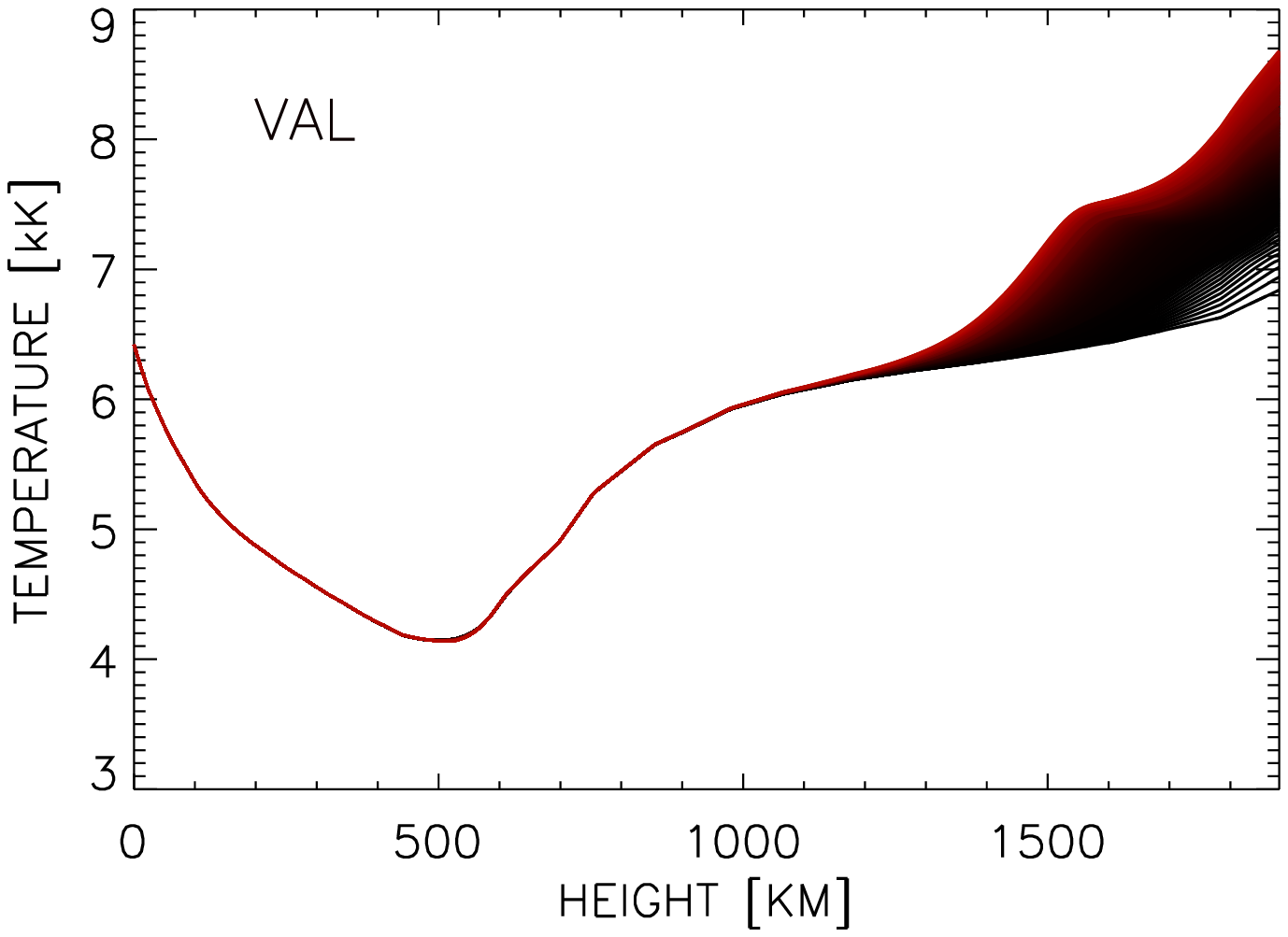}{6cm}{0.}{40.}{40.}{-190}{60}
\plotfiddle{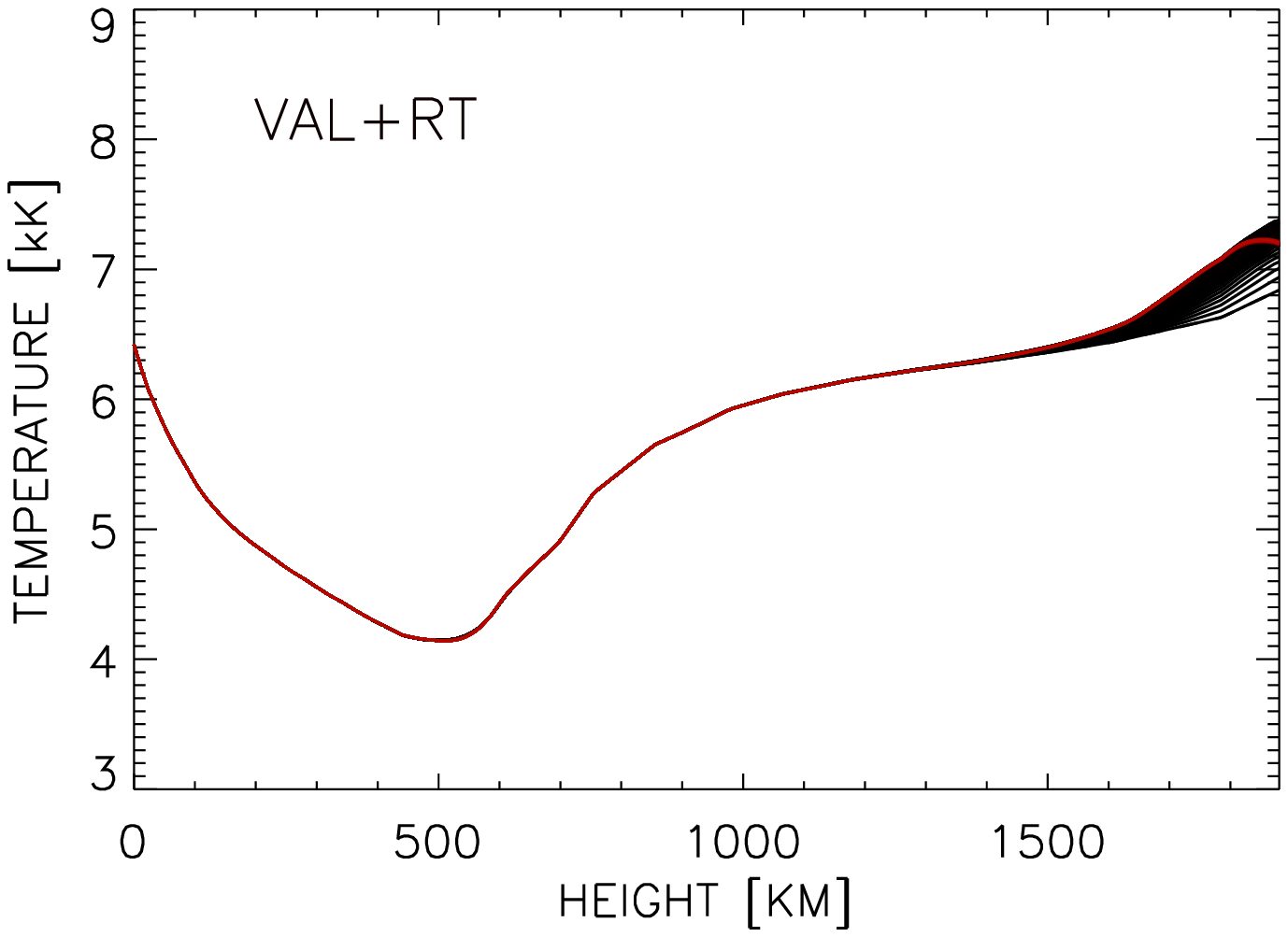}{0cm}{0.}{40.}{40.}{10}{85}
\plotfiddle{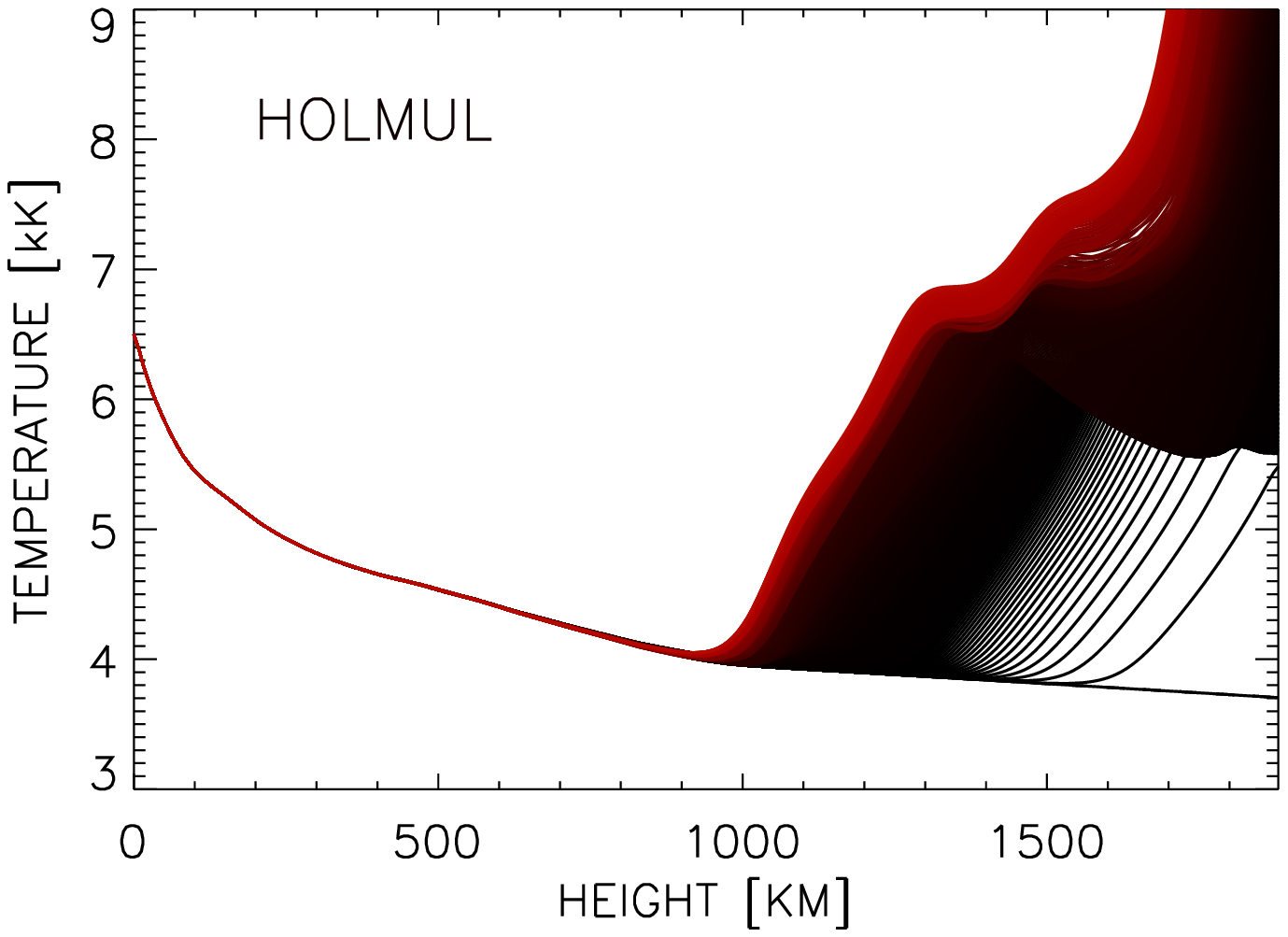}{0cm}{0.}{40.}{40.}{-190}{-30}
\plotfiddle{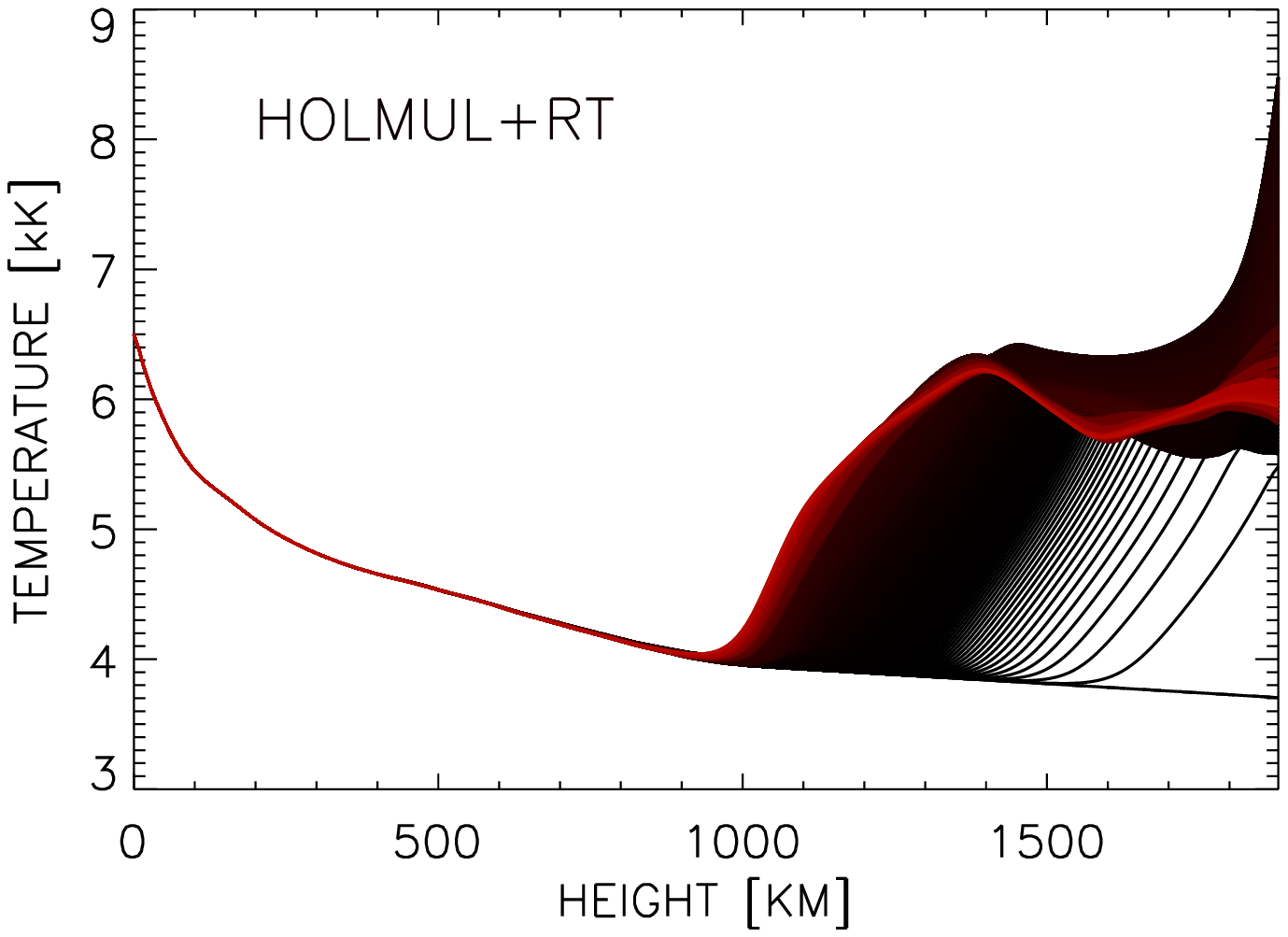}{0cm}{0.}{40.}{40.}{10}{-5}
\caption{{\footnotesize Temperature as a function of height at horizontal
position 0.6 Mm, see Fig. 3. {\it Top left}: VAL-C based flux tube with
$Q_{\rm rad}=0$; {\it top right:} same but $Q_{\rm rad} \neq 0$; {\it bottom
left:} HOLMUL-based flux tube with $Q_{\rm rad}=0$; {\it bottom right:} same,
but $Q_{\rm rad} \neq 0$. Different lines are separated 1 sec in time at the
upper two panels, and 0.5 sec in time at the lower two panels. Progressively
more red colors indicate larger times till 800 sec since the start of the
simulation. }} \label{fig:ft2temp}
\end{figure*}

\begin{figure*}[t]
\center
\plotfiddle{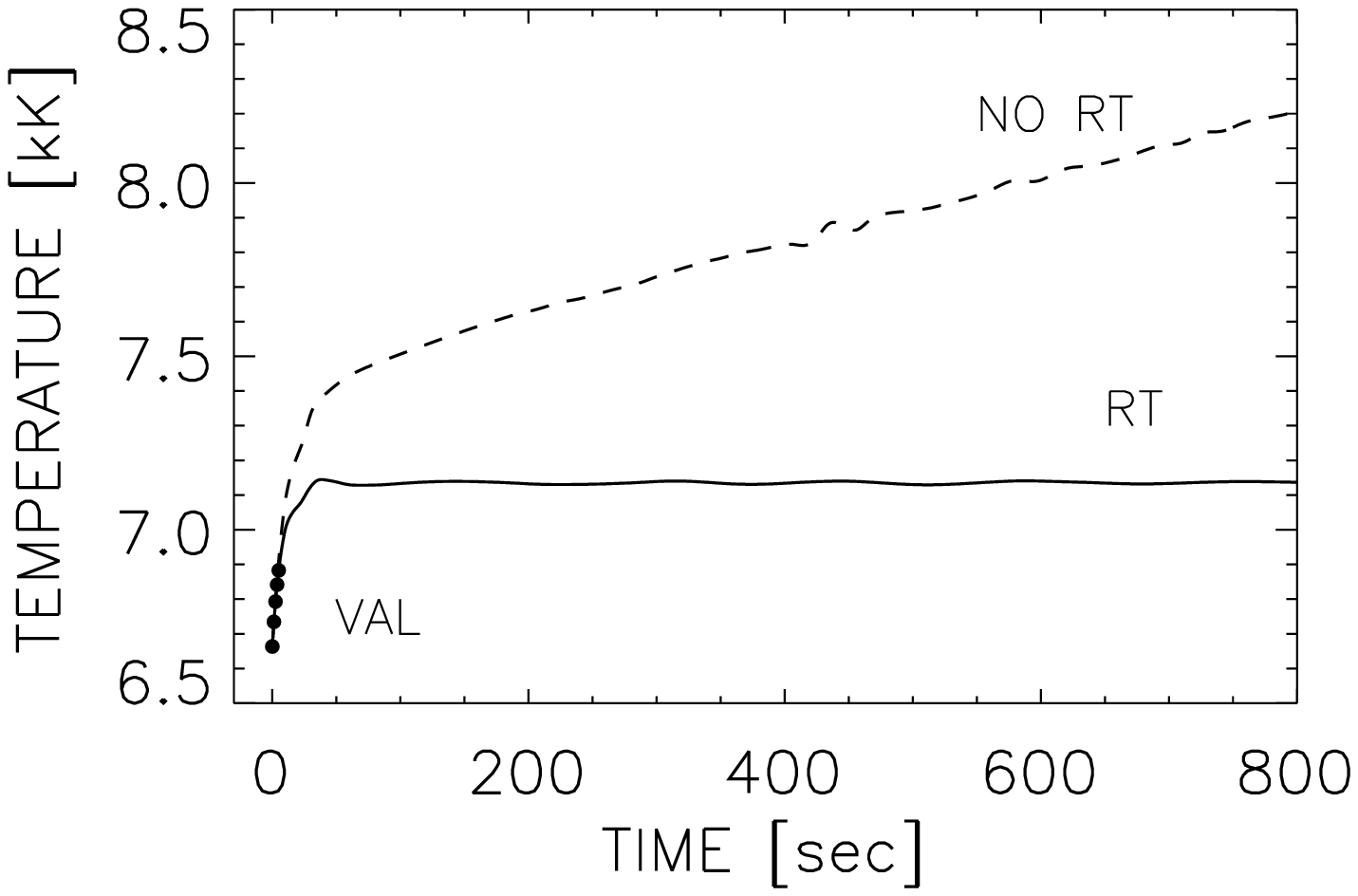}{3cm}{0.}{40.}{40.}{-190}{-30}
\plotfiddle{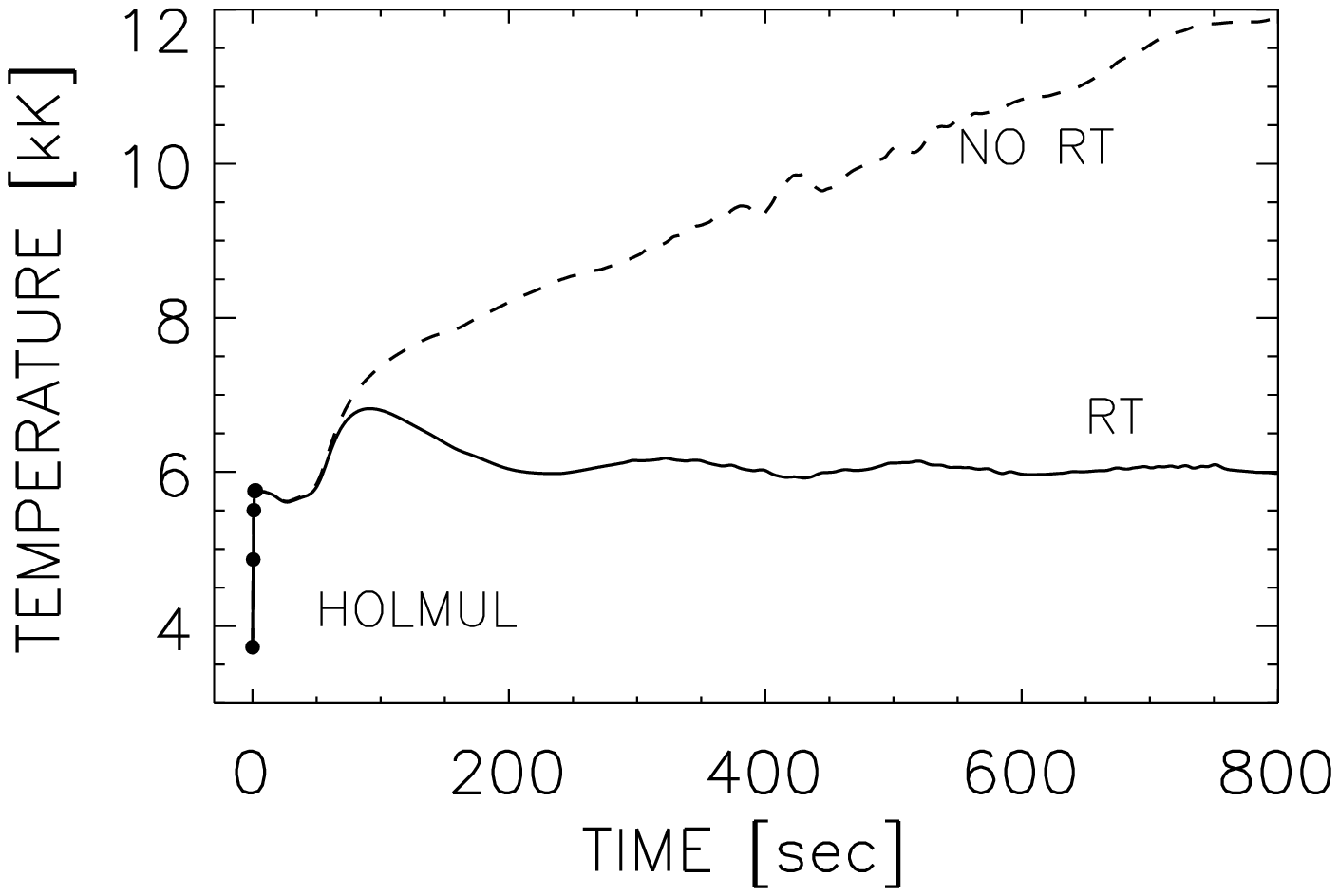}{0cm}{0.}{40.}{40.}{10}{-5}
\caption{{\footnotesize Time variations of temperature at $X=0.6$, $Z=1.8$ Mm
in VAL-based flux tube (left) and HOLMUL-based flux tube (right). Dashed
lines: $Q_{\rm rad}=0$; solid lines: $Q_{\rm rad} \neq 0$. First few time
steps are shown by bullets, separated in time by 1 sec (left) and 0.5 sec
(right).}} \label{fig:ft2tevol}
\end{figure*}

\begin{figure*}[t]
\center
\plotfiddle{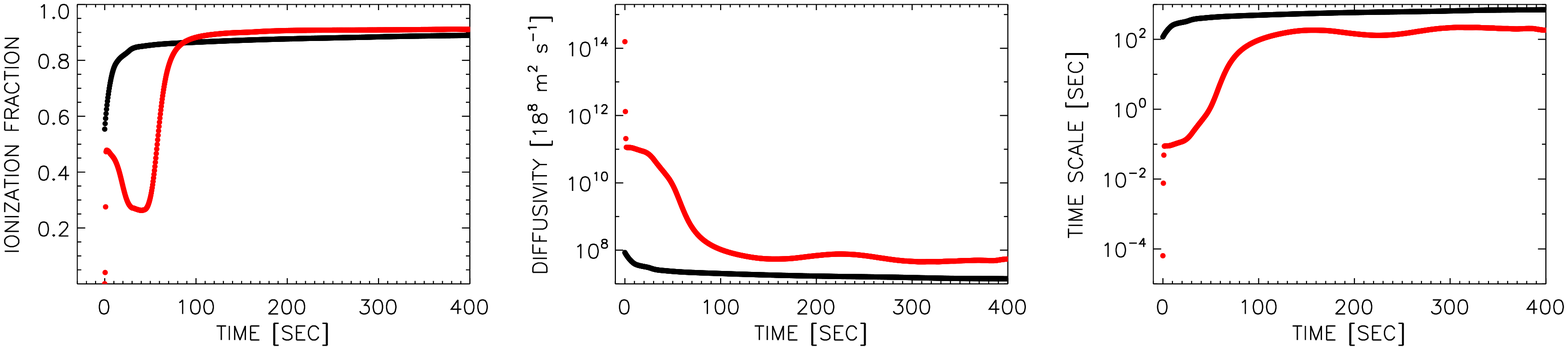}{3cm}{0.}{40.}{40.}{-190}{0}
\caption{{\footnotesize Time variation of the ionization fraction
$\rho_e/\rho$ (left); ambipolar diffusion coefficient $\eta_A$ (middle); and
characteristic time scale $L^2/\eta_A$ (right) in the VAL-based flux tube
(black) and HOLMUL-based flux tube (red), in the simulations with $Q_{\rm
rad}=0$. The values at taken at $X=0.6$, $Z=1.8$ Mm. Bullets are separated
0.5 sec in time.}} \label{fig:ft2ion}
\end{figure*}

\section{Heating of small-scale flux tubes}

Here we describe four simulation runs: (i) VAL-based flux tube with $Q_{\rm
rad}=0$; (ii)  VAL-based flux tube with $Q_{\rm rad}\neq 0$; (iii)
HOLMUL-based flux tube with $Q_{\rm rad}=0$; (iv)  HOLMUL-based flux tube
with $Q_{\rm rad}\neq 0$.
The flux tube models are initially in magneto-static equilibrium, obtained
without considering the diffusion terms. Without external perturbation, they
do not evolve. After introducing the perturbation in the form of diffusion
terms, we perturb the initial magnetic field structure via the induction
equation. Then, as time evolves, this perturbation translates to the rest of
the variables of the system and they start to change. Thanks to the Joule
heating term in the energy equation ($\eta_A\mu_0 J_{\bot}^2$,
Eq.~\ref{eq:system}) the magnetic energy is efficiently converted into
thermal energy, producing heat. This heat is balanced by the radiative
cooling term $Q_{\rm rad}$.

Figure~\ref{fig:ft2evol} shows snapshot from the four simulation runs 800 s
after the introduction of the perturbation. At this time moment, the
temperature has increased at the upper layers of the flux tubes in all
simulations, but by a different amount. A more detailed view of the
temperature behavior is provided in Figures \ref{fig:ft2temp} and
\ref{fig:ft2tevol}. Fig.~\ref{fig:ft2temp} gives the height dependence of the
temperature at a fixed horizontal position inside the flux tube (X=0.6 Mm,
close to flux tubes walls), for different time moments.
Fig.~\ref{fig:ft2tevol} provides the time evolution of the temperature at a fixed
point in the chromosphere, for all four simulations.

In all cases, the most important heating is achieved at the upper part of the
domain, close to the tube borders.  This behavior is expected because the
term responsible for the heating ($\eta_A\mu_0 J_{\bot}^2$) is orders of
magnitude larger at these locations \citep[see Fig. 9 in
][]{Khomenko+Collados2012}.

In the simulations with $Q_{\rm rad}=0$ (left panels of
Figs.~\ref{fig:ft2evol}, \ref{fig:ft2temp}) the relative temperature
increase, achieved after 800 sec, is significantly larger in the cooler
HOLMUL model. In the VAL-based model, the temperature at 1.8 Mm reaches 8200
K after 800 sec of the simulation, which is $\sim$1600 K above its initial
value. In the HOLMUL-based model, it reaches 12000 K, i.e. $\sim$8200 K above
its initial value. This disparity in the amount of heating is readily
understood given the order-of-magnitude different $\eta_A$ values
(Fig.~\ref{fig:ft2initial}). The temperature is visibly enhanced above
$1-1.2$ Mm height. At these heights the action of the ambipolar diffusion
becomes important in our model.

There is a large difference in the time scale of heating between the VAL-base
and HOLMUL-based models. We can roughly define a time scale as $L^2/\eta_A$,
setting a characteristic spatial scale of the system to $L=10^5$ m. Figure
\ref{fig:ft2ion} shows temporal variations of the time scale in the VAL-based
and HOLMUL-based models. It also shows variations of the ionization fraction
and $\eta_A$. In the HOLMUL-based model the characteristic time scale is
initially as low as $10^{-4}$ sec, producing an almost immediate increase of
the temperature from 3700 K to 5700 K (Fig.~\ref{fig:ft2tevol}, bullets on
the right panel). This temperature rise causes an increase of the ionization
fraction from $10^{-3}$ to $\sim0.5$, and a drop of the $\eta_A$ some three
orders of magnitude at the first few seconds. After the initial rapid
variation, the evolution becomes smoother, with characteristic times around
100 sec. Note that VAL-based model does not show such quick changes at the
beginning of the simulation, since $\eta_A$ is initially much smaller than in
the HOLMUL-based model.

While the temperature constantly increases in the simulations with $Q_{\rm
rad}=0$, there is an oscillatory-like balance established in the simulations
with $Q_{\rm rad} \neq 0$ (Fig.~\ref{fig:ft2temp}, right panels;
Fig.~\ref{fig:ft2tevol}, solid lines). After an initial increase, we observe
damped oscillations of temperature converging to some constant value. For
example, at the location $X=0.6$, $Z=1.8$ Mm, the value of temperature to
which the simulations converge is 7100 K (VAL-based, 500 K above the initial
value), and 6000 K (HOLMUL-based, 2300 K above the initial value). Thus,
keeping in mind all the approximations and simplifications of our modeling,
we conclude that the Joule heating and radiative cooling terms in the energy
equation can balance each other.

\section{Conclusions}

We have performed numerical simulations showing that the solar chromosphere
can be effectively heated due to the Joule dissipation of electric currents,
enhanced in the presence of neutral atoms (ambipolar diffusion). Our main
conclusions are:
\begin{itemize}

\item The amount of heating and its time scale depend on the initial
    temperature of chromospheric magnetic structures. In cooler regions
    the heating can act extremely rapidly, reaching a temperature
    increase of 2 kK in few seconds time, while in the hotter regions the
    heating time scale is of the order of minutes.

\item The Joule heating by ambipolar diffusion may be able to balance
    radiative losses of the chromosphere. Our simulations show that after
    a period of damped oscillations, the temperature stabilizes at some
    constant value. In the particular case of the simulations considered
    here this value is $6-7$ kK at 1.8 Mm.

\end{itemize}

\acknowledgements This work is partially supported by the Spanish Ministry of
Science through projects AYA2010-18029 and AYA2011-24808. This work
contributes to the deliverables identified in FP7 European Research Council
grant agreement 277829, ``Magnetic connectivity through the Solar Partially
Ionized Atmosphere'', whose PI is  E. Khomenko (Milestone 4 and contribution
toward Milestone 1).

\bibliographystyle{asp2010}
\bibliography{khomenko}

\end{document}